\documentclass[10pt,letterpaper]{article}
\usepackage[top=0.85in,left=2.75in,footskip=0.75in,marginparwidth=2in]{geometry}

\usepackage[utf8]{inputenc}
\usepackage{float}
\usepackage{cite}

\usepackage{nameref,hyperref}

\usepackage[right]{lineno}
\usepackage{amsmath}
\usepackage{alltt}
\usepackage{braket}
\usepackage{microtype}
\DisableLigatures[f]{encoding = *, family = * }

\raggedright
\usepackage{changepage}

\usepackage[aboveskip=1pt,labelfont=bf,labelsep=period,singlelinecheck=off]{caption}

\makeatletter
\renewcommand{\@biblabel}[1]{\quad#1.}
\makeatother

\usepackage{lastpage,fancyhdr,graphicx}
\usepackage{epstopdf}
\fancyheadoffset[L]{2.25in}
\fancyfootoffset[L]{2.25in}

\usepackage{color}

\definecolor{Gray}{gray}{.25}

\usepackage{graphicx}

\usepackage{sidecap}

\usepackage{wrapfig}
\usepackage[pscoord]{eso-pic}
\usepackage[fulladjust]{marginnote}
\reversemarginpar

\begin{document}
\vspace*{0.35in}

\begin{flushleft}
{\Large
\textbf\newline{High-Energy Neutrino Flavor State Transition Probabilities}
}
\newline
\\
John Harrison\textsuperscript{1,2*}, 
Richard Anantua\textsuperscript{1,3}

\bigskip
\bf{1} Department of Physics and Astronomy, University of Texas at San Antonio, San Antonio, TX 78249, USA
\\
\bf{2} Department of Mathematics and Statistics, Texas A\&M University, Corpus Christi, TX 78412, USA
\\
\bf{3} Department of Physics and Astronomy, Rice University, Houston, TX 77005, USA
\bigskip

*john.harrison@utsa.edu or john.harrison@tamucc.edu

\end{flushleft}
\begin{abstract}
We analytically determine neutrino transitional probabilities and abundance ratios at various distances from the source of creation in several astrophysical contexts, including the Sun, supernovae and cosmic rays. In doing so, we determine the probability of a higher-order transition state from $\nu_\tau\rightarrow\nu_\lambda$, where $\nu_\lambda$ represents a more massive generation than Standard Model neutrinos. We first calculate an approximate cross section for high-energy neutrinos which allows us to formulate comparisons for the oscillation distances of solar, supernova and higher-energy cosmic ray neutrinos. The flavor distributions of the resulting neutrino populations from each source detected at Earth are then compared via fractional density charts.
\newline
$\qquad$
\end{abstract}

$\qquad$
\newline
$\qquad$
\newline
$\qquad$
\newline
\section{Introduction}
\indent 
As neutrinos oscillate 
along their path, their quantum mechanical wave packets develop phase shifts that change how they combine to produce a 
superposition of the three flavors $\nu_e$, $\nu_\mu$ and $\nu_\tau$. 
Thus,  what starts as an electron neutrino may be detected in a detector as a muon or tau neutrino. 
On Earth, the electron neutrino, $\nu_{e}$, 
is the most often detected. This makes sense as this state is what we would generally expect from Solar neutrinos \cite{Borexino(2014)}. 

Let us 
first recognize that  
the flavor states 
can accommodate an ultra-relativistic left-handed neutrino. We then allow this neutrino to have a flavor $\alpha\ (\alpha=e,\mu,\tau)$ and assume the neutrino to have a 4-momentum $P^\mu$ with spatial component, $\vec{p}$. 
Following \cite{Giunti(2007)}, 
we write flavor states of the newly created neutrino
\begin{equation}
\ket{\nu_\alpha}=\Sigma_{k}U^{*}_{\alpha{k}}\ket{\nu_{k}},
\end{equation}
where $U_{\alpha k}$ is a unitary mixing matrix. 


According to Giunti and Kim \cite{Giunti(2007)}, observations 
show 
that our acquired state, $\ket{\nu_{k}}$, is 
a massive neutrino state, 
with its momentum $\vec{p}$ as an eigenstate of the vacuum Hamiltonian, $\mathcal{H}_{0}$, 
which 
can be written 
as

\begin{equation}
\mathcal{H}_{0}\ket{\nu_{k}}=E_{k}\ket{\nu_{k}}
\end{equation}
\ 
where, $E_{k}=\sqrt{\vec{p}^2+m_{k}^{2}}$

\ 

The total Hamiltonian in matter can then be written as

\begin{equation}
\mathcal{H}=\mathcal{H}_{0}+\mathcal{H}_{1}
\end{equation}
where  $\mathcal{H}_{1}\ket{\nu_{\alpha}}=V_{\alpha}\ket{\nu_\alpha}$ \cite{Giunti(2007)}.
This tells us 
that our effective potential from our initial neutrino, $\nu_{\alpha}$, is then described by the component, $V_{\alpha}$ \cite{Giunti(2007)}. We have seen that propagation of neutrinos is of quantum mechanical interest due to the phenomenon of state mixing; the production of neutrinos, on the other hand, garners appeal due to its implications for astrophysical environments.

\par
The interior of a celestial body produces its own magnetic field through currents created by its rotation, which significantly influences the behavior of charged particles on the body's outer surface. When examining a star's collapse, it is essential to consider the acceleration of protons within the core of the collapsing star. This acceleration primarily results from the increasing pressure during the infall phases of the collapse. Additionally, cosmic rays, which are a characteristic of a planet's magnetosphere, may also contribute to the acceleration of particles from the core \cite{Bustamente(2009)}. 
As part of 
this study, we investigate the potential generation of neutrinos from these cosmic rays.

\par
Several elements contribute to understanding the acceleration of particles within a magnetosphere. This study focuses on a few of these elements, specifically the observation and analysis of distant cosmic rays, as well as first-order and second-order Fermi acceleration. Fermi acceleration describes the process by which a charged particle acquires energy due to shock speed or the movement of magnetic mirrors-- with the order indicating how the energy gained relates to velocity. In the first-order process, magnetic fluctuations within shockwaves facilitate the acceleration. An example of first-order Fermi acceleration can be seen during a star's collapse, particularly 
during the bounce phase. Just before the material rebounds after collapsing inward, a shockwave is generated, aiding in the acceleration of charged particles within the magnetosphere itself \cite{Greco(2010)}. Supernova remnant shocks serve as a significant astrophysical environment for first-order Fermi acceleration (diffusive shock acceleration) and are examined here as a potential site for neutrino production \cite{Bustamente(2009)}.
\par During 
second-order Fermi acceleration, 
stochastic acceleration by magnetic mirrors will have a role in answering each of our questions as well as playing a major part in the solution with regard to the probability of a complete flavor transformation. Magnetic mirrors assume the role of bouncing a particle back and forth accelerating it outward. 
The magnetic mirrors are in the form of a magnetic cloud pushing the charged particles along \cite{Greco(2010)}. 

\section{Neutrino Cross Sections}
Since the first direct neutrino detection by Frederick Reines and Clyde Cowan in 1956 \cite{Reines/Cowan(1953)}, neutrino cross sections have been notoriously small-- concordant with the 
weakly interacting nature of neutrino interactions with matter. When considering the cross section as a product of energy gain, $\langle\frac{\Delta{E}}{E}\rangle=\frac{8}{3}(\frac{v}{c})^{2}$, we find the individual interaction cross sections estimated to range from 
$\sigma\equiv\sigma_{\nu_\mu-e\to \mu-\nu_e}\sim 1.33\times 10^{-38}$$\mathrm{cm}^{2}$ 
to $\sim 2.67\times 10^{-38}$$\mathrm{cm}^{2}$
, where $v$ is the speed of the magnetic cloud and $c$ describes the speed of the particle \cite{Bustamente(2009)}.

In a lecture given at Fermilab in 2019 \cite{{Fermilab(2019)}}, it was shown that the simplest method for calculating cross sections relative to the event rate is by finding that 

\begin{equation}
\frac{dN_{r}}{dt}=\sigma_{r}\cdot \frac{dN_{f}}{dt}\cdot n_{b}\cdot d
\end{equation}
where, $N_r$ is the number of reactions, $n_{b}\cdot{d}$ is the number of targets per surface unit with $d$ being the individual target width. Here, $N_{f}$ describes the number of particles within a beam \cite{{Fermilab(2019)}}\cite{{Krauss(2019)}}. 
After integrating over $t$, we find 

\begin{equation}
\sigma_{r}=\frac{N_{r}}{N_{f}}\frac{1}{n_{b}\cdot{d}}
\end{equation}

\begin{figure}[h!]
\centering
\includegraphics[scale=0.4]{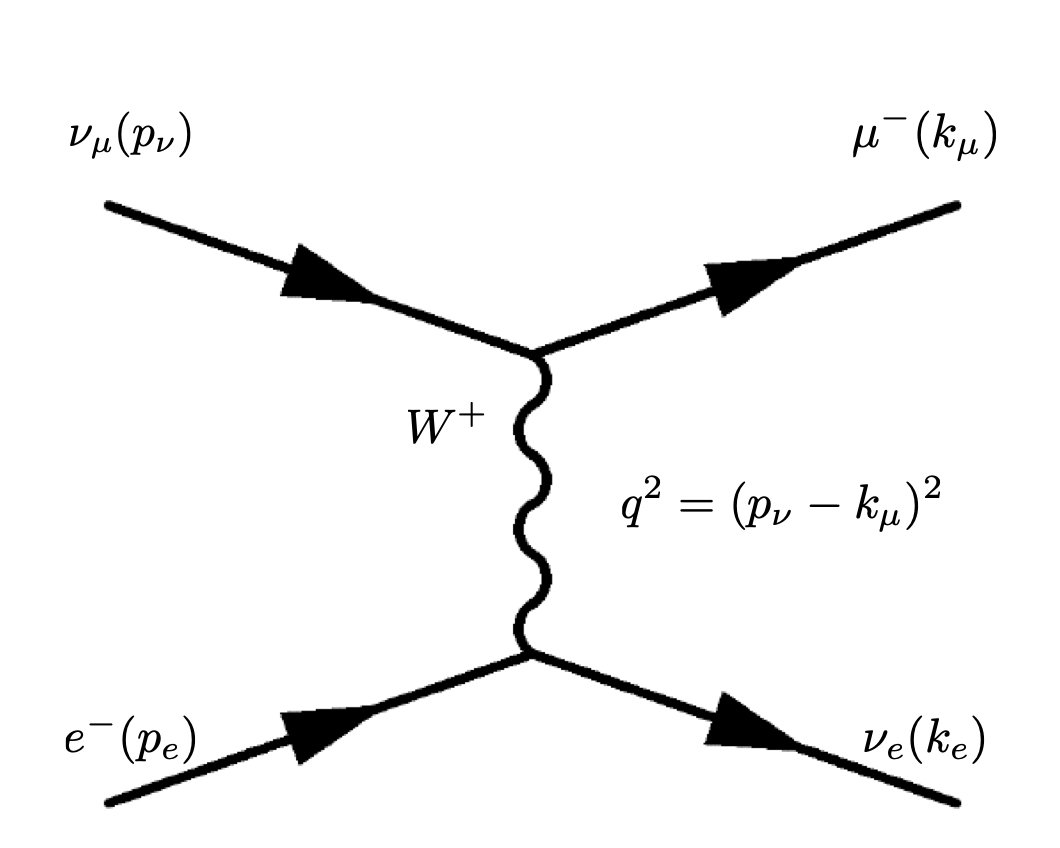
}
\caption{Feynman diagram depicting the scattering of an incoming muon-neutrino and an at rest electron. Figure used with permission of author \cite{Formaggio(2013)}.}
\label{fig: FEYNMAN}
\end{figure}

The cross section for the interaction illustrated in Figure \ref{fig: FEYNMAN}, in which a muon neutrino interacts with an electron to produce an electron neutrino and a muon, can be succinctly expressed as a function of neutrino energy by,

\begin{equation}
\sigma\approx \frac{2m_{e}G_{F}^{2}E_{\nu}}{\pi}=\frac{G_{F}^{2}s}{\pi}
\end{equation}

However, to compute cross sections at high, or ultra-high energies such as that of cosmic rays, we have that 

\begin{equation}
\begin{split}
\frac{d\sigma(\nu_{e})}{dy}=\frac{m_{e}G_{F}^{2}E_{\nu}}{2\pi}((g_{V}-g_{A})^{2}+\\
(g_{V}-g_{A}^{2})(1-y)^{2}-(g_{V}^{2}-g_{A}^{2})
(\frac{m_{e}y}{E_{\nu}}))
\end{split}
\end{equation}

where, $G_{F}=\frac{g^{2}}{4\sqrt{2}M^{2}_{W}}\approx 1.166\times 10^{-5}\: \mathrm{GeV}^{-2}$, $y=2m_{e}E_{\nu}$ which is the outgoing leptonic energy, bounded by $0\leq y \leq y_{max}=1-\frac{m_{l}^{2}}{2m_{e}E_{\nu}+m_{e}^{2}}$. The coupling constants are given as $g_{L}=+1/2, g_{R}=0, g_{V}=+1/2$ and $g_{A}=+1/2$ in which case $L, R, V, A$ are the left, right, vector and axial components, respectively \cite{Formaggio(2013)}.
\par
We can then determine approximate cross sections for high and ultra high energy cosmic rays. First, by substituting in our components $m_{e}=9.11\times10^{-31}\: \mathrm{kg}$, with an energy of $300\: \mathrm{GeV}$ we obtain a value for $y$ of $\sim5.466\times10^{-28}$ which gives an approximate cross section of $5.072\times10^{-34}\: \mathrm{cm}^{2}$. Next, we can compare that with an energy level of $500\: \mathrm{GeV}$ or $0.5\: \mathrm{TeV}$. This gives us a $y$ value of $\sim9.11\times10^{-28}$ in which we deduce a cross section of approximately $8.453\times10^{-34}\:\mathrm{cm}^{2}$. Lastly we will use an energy of $E_{\nu}=1,000\: \mathrm{GeV}$ or $1\:  \mathrm{TeV}$, which is the energy level used in determining our cosmic ray probability calculations. The component $y$ we find to be on the order of $\sim1.822\times10^{-27}$. This yields an approximate neutrino cross section for cosmic rays as approximately $1.691\times 10^{-33}\: \mathrm{cm}^{2}$. 
 
\section{Oscillation and the Mean Free Path}

Using 
the 
conversion probability function, we can ascertain the travel distance for oscillation and anticipate the state of our neutrino at a specific position and time. To simplify this process, it helps to conceptualize oscillating neutrinos as wave packets existing in some arbitrary state,
$\ket{\psi}$. This arbitrary state is our initial condition in which we have $\ket{\psi_{\alpha}}$ where, $\alpha$ is the flavor eigenstate pre-measurement (or detection), $\ket{\nu_{e}, \nu_{\mu}, \nu_{\tau}}$. Similarly, 
$\ket{\psi_{\alpha}'}$ 
is our final state after detection. 
We must also consider the probability of measurement of the mass eigenstate. For the sake of simplification, we 
use the definition for neutrino flavor states as explained by Blasone, et al. (2022) \cite{Blasone(2022)}, and utilizing the $3\times 3$ Pontecorvo-Maki-Nakagawa-Sakata [PMNS] unitary mixing matrix \cite{Koranga(2020)}, $\textbf{U}(\tilde{\theta},\delta)$, as follows:

\begin{equation}
\ket{\underline{\nu}^{(f)}}=\textbf{U}({\tilde{\theta},\delta})\ket{\underline{\nu}^{(m)}}
\end{equation}

\par
It is imperative to note, that in the above equation, we must explicitly show that there are two possible eigenstates: $\ket{\underline{\nu}^{(f)}}$ is our flavor eigenstate representing the transpose of $(\ket{\nu_{e}}, \ket{\nu_{\mu}},\ket{ \nu_{\tau}})$. Then we have that $\ket{\underline{\nu}^{(m)}}$ is our mass eigenstate, which is represented by the transpose of $(\ket{\nu_{1}}, \ket{\nu_{2}},\ket{ \nu_{3}})$. The probability for the flavor transition $\nu_{\alpha}\rightarrow{\nu_{\beta}}$ \cite{Giunti(2007)} can be defined as 

\begin{equation}
P_{\nu_{\alpha}\rightarrow{\nu_{\beta}}}(t)=|\bra{\nu_{\beta}}\ket{\nu_{\alpha}(t)}|^{2}=|\tilde{\textbf{U}}_{\alpha\beta}(t)|^{2}
\end{equation}
This indicates the flavor transition at time, $t=0$ with the contribution from \cite{Blasone(2022)}. 
Here we can assume from earlier that our variables, $\alpha$ and $\beta$, can represent 
any of the flavors $(e,\mu,\tau)$ and we have that $\tilde{\textbf{U}}(t=0)$ is normalized \cite{Blasone(2022)}.
\par
We know from 
above that our neutrino flavor mixing matrix can be 
written as 

\begin{equation}
\begin{pmatrix}
\nu_{e}  \\
\nu_{\mu}  
\end{pmatrix}=  
\begin{pmatrix}
\cos\theta & \sin\theta \\
\mathrm{-sin}\:\theta & \cos\theta 
\end{pmatrix}
\begin{pmatrix}
\nu_{1}  \\
\nu_{2}  
\end{pmatrix}
\end{equation}

We can now show this as a unitary mixing matrix 

\begin{equation}
\textbf{U}(\theta)=\begin{pmatrix}\cos\theta & \sin\theta \\
\mathrm{-sin}\:\theta & \cos\theta 
\end{pmatrix}
\end{equation}

Assuming that neutrinos have the ability to oscillate into three, four, or even an infinite number of flavors, we can illustrate the mixing phenomenon by examining a state involving two flavors. It is important to note that in the previous discussion, neutrinos were regarded as flavor states rather than as a combination of mass eigenstates,
$\nu_{1},\nu_{2},\nu_{3}$. With this in mind, we can consider two weak eigenstates, $\nu_{e}$ and $\nu_{\mu}$. Their plane wave forms \cite{Giunti(2007)} can be expressed as

\begin{equation}
\ket{\nu_{1}(t)}=\ket{\nu_{1}}e^{i(\textbf{p}_{1}\cdot x-E_{1}t)}=\ket{\nu_{1}}e^{-ip_{1}\cdot x}
\end{equation}

and also

\begin{equation}
\ket{\nu_{2}(t)}=\ket{\nu_{2}}e^{i(\textbf{p}_{2}\cdot x-E_{2}t)}=\ket{\nu_{2}}e^{-ip_{2}\cdot x}
\end{equation}

If we let $A$ be the flavor eigenstate consisting of $e,\mu,\tau$ 
and let $B$ be the neutrino mass eigenstate consisting of $m_{1},m_{2}, m_{3}$ then we must consider that $A\times{B}\neq{B}\times{A}$. Assuming this constraint we can deduce that \\
$A\times{B}=\{(e,m_{1}),(e,m_{2}),(e,m_{3}),(\mu,m_{1}),(\mu,m_{2}),(\mu,m_{3}),
(\tau,m_{1}), (\tau,m_{2}), (\tau,m_{3})\}$.
Similarly, \\ 
$B\times A $ $=\{(m_1,e),(m_2,e),$ $(m_3,e),(m_1,\mu),(m_2,\mu),(m_3,\mu),
(m_1,\tau),(m_2,\tau),(m_3,\tau)\}$
. Again we must stress the importance that for this assumption, the probability of $A$ does not 
depend on the probability of $B$ and the probability of $B$ certainly does not depend on the probability of $A$. 
In other words, if we determine a state in $A$, that does not mean we know for certain state $B$, rather 
that all other probabilities for state $B$ go to zero. 
\par
To consider the probability of an arbitrary neutrino state prediction, we must first consider a known constraint. To make a prediction based on a flavor eigenstate, we must assume that all mass eigenstate probabilities collapse. Conversely, if we are to make a prediction based on a mass eigenstate, all the flavor eigenstate probabilities collapse. 
\par
However, we must consider our neutrino eigenstates: $\ket{\nu_e,\nu_\mu,\nu_\tau}$ and $\ket{m_{1},m_{2},m_{3}}$. Because we are aware of our constraints, the same assumption made previously strictly applies here. That is the outcome observed for $A$ does not depend on the outcome observed for $B$ and the outcome observed for $B$ does not depend on the outcome observed for $A$. 
\par
We can now determine the probability for flavor transition from $\nu_{e}\rightarrow\nu_{\mu}$ which corresponds to the detection of $\nu_{e}$ and is dependent on $\Delta{m}$ as the mass difference, where, $\Delta{m}_{ij}^{2}=m_{i}^{2}-m_{j}^{2}$, $E$ as the neutrino energy and the distance from the source, $L$.

\begin{equation}
P(\nu_{e}\rightarrow\nu_{\mu})=1-sin^{2}(2\theta_{21}) sin^{2}( \frac{1.27\Delta{m_{21}^{2}[eV]L[m]}}{E[MeV]} )
\end{equation}

Alternatively, we can show the probability of our mass eigenstate and the prediction of the flavor state remaining in $\nu_{e}$. We  will reserve component, $\lambda$, as an arbitrary flavor state. The constraint here, however, is certainty in one eigenstate negates certainty in the other \cite{Giunti(2007)}. 
\par
For comparative purposes, on Earth, we can expect a solar neutrino flux on the order of $\sim5.9\times10^{14}\ m^{-2}$ with approximately $1.7\times10^{38}$ neutrinos being emitted every second. Considering the Type II supernova, SN1987A, located in the Large Magellanic Cloud, which was first detected on February 23, 1987. During its luminosity peak, approximately $5.9\times10^{19}$ neutrinos were detected on Earth per second with an average flux of approximately $1.3\times10^{14} \mathrm{m}^{-2}$. The distance $d$ from SN1987A to Earth is $\sim168,000$ light years \cite{Page(2020)}. 

\section{
Sterile Neutrinos}
As we consider the neutrino's ability to oscillate 
among $\nu_e,\nu_\mu,\nu_\tau$,
it is theoretically possible to oscillate into a heavier fourth flavor, without undergoing further oscillations, provided that this flavor interacts solely through the gravitational force. 
Due to this property, this fourth flavor has received the name sterile neutrino. It is essential to recognize that the sterile neutrino functions independently from the three active flavors. When a neutrino is generated, it can be described as a superposition of the three active flavors, each associated with their respective masses, while the sterile neutrino remains excluded from this interaction. Like other neutrinos, the sterile neutrino is electrically neutral and possesses a half-integer spin. As noted, this type of neutrino is anticipated to be heavier than the three known flavors, with theoretical mass estimates ranging from $10$ GeV and $1$ eV \cite{Akhmedov(2006)}.

\par
When we consider the existence of a sterile neutrino, we must also consider the probability of it being categorized as a Majorana particle. A fermion is considered Majorana when it is found to be its own antiparticle. Typical fermions are categorized as Dirac fermions at low energies, with the exception of the neutrino. Dirac fermions do not possess their own antiparticles. At this point, which category the sterile neutrino fits is unknown. If sterile neutrinos are Majorana, when two sterile neutrinos collide, they would annihilate each other with an enormous amount of energy. 
\par 
We 
classify these particles as fermions, i. e., with a half-integer spin. In the context of neutrinos, 
this means their spin may either be $1/2$  or $-1/2$ depending on the spin quantum number $s$ quantifying direction of intrinsic rotation. 
Neutrinos are typically considered to be ``left-handed" fermions, 
i.e., with Dirac spinor representations that can be expressed as projections by the left-handed projection operator $P_L=\frac{1-\gamma^5}{2}$. 
Moreover, 
the $W^\pm$ bosons mediating weak interactions involving neutrinos only interact with left-handed particles. When considering sterile neutrinos, they are assumed to be ``right-handed" fermions, thus confirming the existence of a positive-spin neutrino.  	
\par 
If we take the mass of one ``hand," multiplied by the other ``hand" we would deduce a constant mass. 

So we can show that:

\begin{equation}
(Right Hand Mass)\times(Left Hand Mass)=Constant    
\end{equation}

Therefore, we can now assume that if we were to increase the mass of one hand, that is, left or right, the other hand must become smaller. This results in a process referred to as the \textit{``Seesaw Mechanism"} \cite{Akhmedov(2006)}

\begin{equation}
M_{(\nu)}=\begin{pmatrix}\nu_{L}f & \nu y \\
\nu y & \nu_{R}f
\end{pmatrix}
\end{equation}
where, $f$ directly corresponds to neutrinos of heavier mass, however, a mixture of $f$ results in neutrinos of lighter mass \cite{Akhmedov(2006)}. This yields the mass eigenstate of

\begin{equation}
M_{(\nu)}=\begin{pmatrix}0 & M_{D} \\
M_{D} & M_{NHL}
\end{pmatrix}
\end{equation}
in which the mass hierarchy can be shown as $M_{\nu}>> M_{D}>> M_{NHL}$, where $M_{\nu}$ is the neutrino mass, $M_{D}$ is the mass of the corresponding deuteron and $M_{NHL}$ represents the neutral-heavy lepton \cite{Akhmedov(2006)}. By the Seesaw Mechanism, this yields an approximate neutrino mass of 

\begin{equation}
M_{\nu}=\frac{M_{D}^{2}}{M_{NHL}}
\end{equation}

\par
Once again, the Seesaw Mechanism shows that if these right-handed sterile neutrinos exist, they could in fact be categorized as Majorana, which could help explain the difficulty in detecting them. This mechanism gives us a matrix theorizing six neutrino fields, three of which would have masses $<1\:\mathrm{eV}$ while the other three would have masses $ >1\:\mathrm{eV}$. Unlike other fermions, these Majorana particles do not exhibit intrinsic electric or magnetic moments. Instead, they would possess toroidal moments in which the field of the solenoid is bent into a torus. 

\section{Methods 
}
We have previously shown the high probability of detecting our neutrino to be in states, $\nu_{e}$ or $\nu_{\mu}$. To consider a higher flavor transition, we must first consider the transition from $\nu_{\mu}$ to $\nu_{\tau}$ whose probability can be shown as \cite{Giunti(2007)} 

\begin{equation}
P(\nu_{\mu}\rightarrow\nu_{\tau})=\sin^{2}(2\theta_{32}) \sin^{2}\left( \frac{1.27\Delta{m_{32}^{2}[eV]L[m]}}{E[MeV]} \right)
\end{equation}

We now need to expand the PMNS matrix \cite{Koranga(2020)} to include a fourth flavor by considering that:

\begin{equation}
\begin{split}
|m_{ee}|_{4\nu}=|\Sigma|U_{ej}|^{2}e^{i\phi_{j}}m_{j}|=\\
|c_{13}^{2}c_{12}^{2}c_{14}^{2}m_{1} 
  +c_{13}^{2}c_{12}^{2}c_{14}^{2}e^{i\alpha}m_{2}\\
+ s_{13}^{2}c_{14}^{2}e^{i\beta}m_{3} 
  +s_{14}^{2}e^{i\gamma}m_{4} 
\end{split}
\end{equation}

\begin{equation}
\begin{split}
m_{ee}=|\Sigma|U_{ej}|^{2}e^{i\phi_{j}}m_{j}|=\\
|c_{13}^{2}c_{12}^{2}c_{14}^{2}m_{1}
  +c_{13}^{2}c_{12}^{2}c_{14}^{2}e^{i\alpha}\sqrt{m_{1}^{2}+\Delta_{21}}\\
+ s_{13}^{2}c_{14}^{2}e^{i\beta}\sqrt{m_{1}^{2}+\Delta_{31}}
  +s_{14}^{2}e^{i\gamma}\sqrt{m_{1}^{2}+\Delta_{41}} 
\end{split}
\end{equation}

\begin{equation}
\begin{pmatrix} \nu_{e} \\
\nu_{\mu} \\
\nu_{\tau} \\
\nu_{\lambda} \\
\end{pmatrix}=
\begin{pmatrix}U_{e1} & U_{e2} & U_{e3} & U_{e4} \\
U_{\mu1} & U_{\mu2} & U_{\mu3} & U_{\mu4} \\
U_{\tau1} & U_{\tau2} & U_{\tau3} & U_{\tau4} \\
U_{\lambda1} & U_{\lambda2} & U_{\lambda3} & U_{\lambda4} 
\end{pmatrix}\begin{pmatrix} \nu_{1} \\
\nu_{2} \\
\nu_{3} \\
\nu_{4} \\
\end{pmatrix}
\end{equation}

where $U=R_{34}R_{24}R_{14}R_{23}R_{13}R_{12}P$ and so that \cite{Koranga(2020)} ,

\begin{equation}
\begin{split}
R_{14}=
\left( \begin{array}{cccccc}
  c_{14} & 0 & 0 &  -s_{14}e^{i\delta_{14}}  \\
  0 & 1 & 0 & 0 \\
  0 & 0 & 1 & 0 \\
  -s_{14}e^{i\delta_{14}} & 0 & 0 & c_{14}
\end{array} \right)
\:\mathrm{and} \\
R_{34}=
\left( \begin{array}{cccc}
  1 & 0 & 0 & 0 \\
  0 & 1 & 0 & 0  \\
  0 & 0 & c_{34} & s_{34}  \\
  0 & 0 & -s_{34} & c_{34} 
\end{array} \right)
\end{split}
\end{equation}

\par
We can now define the probability for a higher-order transition as

\begin{equation}
P(\nu_{\tau}\rightarrow\nu_{\lambda})=\sin^{2}(2\theta_{43}) \sin^{2}\left( \frac{1.27\Delta{m_{43}^{2}[eV]L[m]}}{E[MeV]} \right)
\end{equation}
where, $L[m]$ is the distance from the source, $E[MeV]$ is the energy and $\Delta{m}$ is the mass difference between the flavors during the oscillation. This indicates that the probability of finding a neutrino in the fourth flavor state should be equal to the probability of finding it in the third flavor state. 
\par
Taking what we know from our probability functions, Equations $(12)$ and $(19)$, we can now use this to determine the probability for an $n^{th}$ flavor. If we assume an oscillation into a fourth flavor, the sterile neutrino, is probable, then we can assume an equal probability of continuous oscillation into $n$ flavors. We can now determine the theoretical travel distance for oscillation from $\nu_{e}\rightarrow\nu_{\mu}$ by assuming our mixing angle $\theta$ to be $\sim0.846$, our $|\Delta{m_{21}}|^{2}$ to be approximately $7.4\times10^{-5}\ \mathrm{(eV/c^2)}^2$ and we can assume an approximate energy $E$ of $10^{8}\: \mathrm{eV}$ assuming a supernova neutrino creation. We can show a travel distance of approximately  $10^{21}\: \mathrm{m}$ or around $105,700$ light years, with a probability of $\sim1$ or 
$\sim100\%$ of the transition from electron neutrino, $\nu_{e}$, to muon neutrino, $\nu_{\mu}$. 
Given this amount of travel distance, the neutrino would have definitively transitioned from its electron state, to its muon state.
\par
To ascertain the pattern of transitional probabilities, a Python 3.13 program was developed  to compare transitions between three widely detected neutrino types: solar, supernova and cosmic rays. For our first simulation, the following parameters were used for solar neutrinos:

For the first flavor change we will use for our mixing angle, $\sin^{2}\theta\approx 0.846$ and a mass difference of $\Delta m_{21}^{2}\approx 7.40\times10^{-5}\: \mathrm{eV}^{2}$. For our second flavor change, $\sin^{2}\theta\approx 0.92$ and $\Delta m_{32}^{2}\approx 2.40\times10^{-3}\: \mathrm{eV}^{2}$. For our third flavor change, $\sin^{2}\theta\approx 0.994$ and $\Delta m_{43}^{2}\approx 0.05\: \mathrm{eV}^{2}$.
With an adopted solar neutrino energy of $E[MeV]\approx 1\: \mathrm{MeV}$. \\ \indent

\section{
Results}

We are now in a position to use our expression for neutrino  transitional probabilities to estimate the neutrino transition fractions (among 3 species) at various distances. 
First we turn to our most important neutrino source, the Sun. The transitional probabilities from $\nu_e\to\nu_\mu$ and $\nu_\mu\to\nu_\tau$ are shown  
in Table 1 for solar neutrinos.

\begin{table}[h]

\caption{Oscillation probabilities from Solar neutrinos at various distances from the source of neutrino production.}\label{tab:table1}
\footnotesize
\centering 

\begin{tabular}{@{} llll @{}}
\hline
\textbf{Distance $L[m]$}	& \textbf{P$(\nu_{e}\rightarrow\nu_{\mu})$}	& \textbf{P$(\nu_{\mu}\rightarrow\nu_{\tau})$} & \textbf{P($\nu_{\tau}\rightarrow\nu_{\lambda})$}  \\
\hline
$10^{5}$   & $0.999$    &$0.004$&   $0.464$\\
$10^{10}$   & $0.172$    &$0.691$&   $0.479$\\
$10^{15}$ &   $0.041$    & $0.304$   &$0.528$\\
$10^{20}$  &$0.449$   & $0.650$ & $0.457$\\
$10^{25}$     &$0.677$   & $0.202$ &  $0.570$\\
$10^{27}$     &$0.612$   & $0.591$ &  $0.218$\\
\hline
\end{tabular}

\end{table}

For Table 2, we maintain the parameters used above and apply them to neutrinos detected from supernova events.
We assume 
a median supernova neutrino energy level of $
\sim 10\: \mathrm{MeV}$.

\begin{table}[h]

\caption{Oscillation probabilities from supernova neutrinos at various distances from the source of neutrino production.}\label{tab:table2} 
\footnotesize
\centering 

\begin{tabular}{@{} llll @{}}
\hline
\textbf{Distance $L[m]$}	& \textbf{P$(\nu_{e}\rightarrow\nu_{\mu})$}	& \textbf{P$(\nu_{\mu}\rightarrow\nu_{\tau})$} & \textbf{P($\nu_{\tau}\rightarrow\nu_{\lambda})$}  \\
\hline
$10^{5}$   & $0.358$    &$0.602$&   $0.140$\\
$10^{10}$   & $0.548$    &$0.919$&   $0.399$\\
$10^{15}$ &   $0.205$    & $0.268$   &$0.041$\\
$10^{20}$  &$0.220$   & $0.692$ & $0.792$\\
$10^{25}$     &$0.868$   & $0.117$ &  $0.355$\\
$10^{27}$     &$0.595$   & $0.117$ &  $0.822$\\
\hline
\end{tabular}

\end{table}

Finally, in Table 3, we consider a significant increase in energy. This is partly due to the first- and second- order Fermi processes within the magnetized clouds accelerating the cosmic rays. With this in mind, and again utilizing our base parameters, we now have an increased cosmic ray energy of $\sim 1\times 10^{6}\: \mathrm{MeV}$.

\begin{table}[h]

\caption{Oscillation probabilities from cosmic ray neutrinos at various distances from the source of neutrino production.}\label{tab:table3} 
\footnotesize
\centering 

\begin{tabular}{@{} llll @{}}
\hline
\textbf{Distance $L[m]$}	& \textbf{P$(\nu_{e}\rightarrow\nu_{\mu})$}	& \textbf{P$(\nu_{\mu}\rightarrow\nu_{\tau})$} & \textbf{P($\nu_{\tau}\rightarrow\nu_{\lambda})$}  \\
\hline
$10^{5}$   & $0.999$    &$0.000$&   $0.000$\\
$10^{10}$   & $0.358$    &$0.602$&   $0.126$\\
$10^{15}$ &   $0.548$    & $0.919$   &$0.359$\\
$10^{20}$  &$0.205$   & $0.268$ & $0.041$\\
$10^{25}$     &$0.216$   & $0.692$ &  $0.792$\\
$10^{27}$     &$0.605$   & $0.198$ &  $0.667$\\
\hline
\end{tabular}

\end{table}

From the above tables, we have shown 
the distance needed to oscillate from the first flavor to the second flavor, with a consistently high probability of transition. Once we move into the second, third, and $n^\mathrm{th}$ flavors, we become less certain about our flavor transition probabilities. From the above two tables, we can infer 
why this is the case. We can clearly see that our probabilities for the higher flavors $(\mu\rightarrow\tau, \tau\rightarrow\lambda)$ are the lowest when the probability for the transition from the first to the second flavor $(e\rightarrow\mu)$ is at its highest. For example, for a cosmic ray neutrino, using our transition from $\nu_{\tau}\rightarrow\nu_{\lambda}$ at distances $10^{5}\: \mathrm{m}$ and $10^{10}\: \mathrm{m}$ we obtained probabilities of $\sim0.00\%$ and $\sim12.6\%$, respectively. 
\par
When comparing this with our probability from $\nu_{e}\rightarrow\nu_{\mu}$ at the same distances, we find a probability of $99.9\%$ but a significant decease to $\sim35.8\%$ when $\nu_\mu\rightarrow\nu_\tau$ is $\sim60.2\%$.  However, in Tables 1 and 3 there are distances from which our first flavor transition is approximately equal to our third flavor transition. In the case of Solar neutrinos, at $L[m]\sim10^{20}\: \mathrm{m}$ the probabilities for first and third transitions are $\sim44.9\%$ and $\sim45.7\%$, respectively. For the case with cosmic rays, we have that at $L[m]\sim10^{27}\: \mathrm{m}$ our first transition is $\sim60.5\%$ while our third transition is $\sim66.7\%$. 
\par
If we consider our distance of travel to be on the order of $1$ AU Unit (AU), we find that our probability would be $\sim0.261$ for flavor oscillation into the second state and $\sim0.111$ for higher-order flavor transitions. When we compare this to known detection rates, we find approximately $\frac{2}{3}$ of detected neutrinos from the Sun to found in either $\nu_e$ or $\nu_\mu$ flavor state, which is consistent with our findings in the range of $\sim26.1\%$ found in the $\nu_e$ flavor state \cite{Kamiokande(2024)}. 
\par
A similar assumption can be made for atmospheric neutrinos. As cosmic rays bombard the atmosphere of Earth, protons 
strike the nuclei of atoms within our atmosphere forming pions. These short-lived particles decay into muons which then leave behind an electron, anti-electron and a neutrino, the $\nu_\mu$. This flavor, according to Fermilab, make up approximately $\frac{2}{3}$ of the detected neutrinos on Earth leaving behind $\frac{1}{3}$ to be detected as $\nu_{e}$ \cite{Kamiokande(2024)}. When comparing this with our calculations, we find that with cosmic rays we expect at a distance of $1.5\times10^{11}\: \mathrm{m}$, $\sim{0.399}$ to be in flavor state $\nu_{e}$, while we have $\sim0.813$ to be in state, $\nu_\mu$. 

\par
Taking these transition probabilities into account, we can quickly see that from varying distances and probabilities, the neutrino oscillation of flavors is simply a cycle of probabilities. In other words, the flavor transition can start again depending on the distance traveled prior to detection. What makes this concept 
more 
intriguing 
is that it helps explain a possible reasoning behind only having certainty in one eigenstate and not in both. When we measure a neutrino in a flavor eigenstate, we can assume that, and as the tables above show, as it oscillates between flavors it becomes a combination of all possible masses $m_{1}, m_{2}, m_{3},...,m_{n}$. When we consider the combination of mass matrices, this also helps to understand the cycle of flavor oscillations prior to detection. 
\par
If we are to assume that when we observe a neutrino in state $A$ or $\ket{\nu^{f}}$ we have the probability for $\ket{\nu_{e},\nu_{\mu},\nu_{\tau},\nu_{\lambda}}$ with state $B$ or $\ket{\nu^{m}}$ for $\ket{m_1,m_2,m_3,m_{\lambda}}$. We can predict our condition to be approximately as observed $\nu_{e}$ to have state $B$ of $\ket{m_1,0,0,0}$, for $\nu_{\mu}$ we find $\ket{m_1,m_2,0,0}$, for $\nu_{\tau}$ we would show $\ket{m_1,m_2,m_3,0}$ and for $\nu_{\lambda}$, $\ket{m_1,m_2,m_3,m_{\lambda}}$. This shows that each potential flavor state can be thought of simply as a combination of all possible masses.

\section{Analysis of Results}

\begin{figure}[h]
\centering
\includegraphics[width=0.45\textwidth]{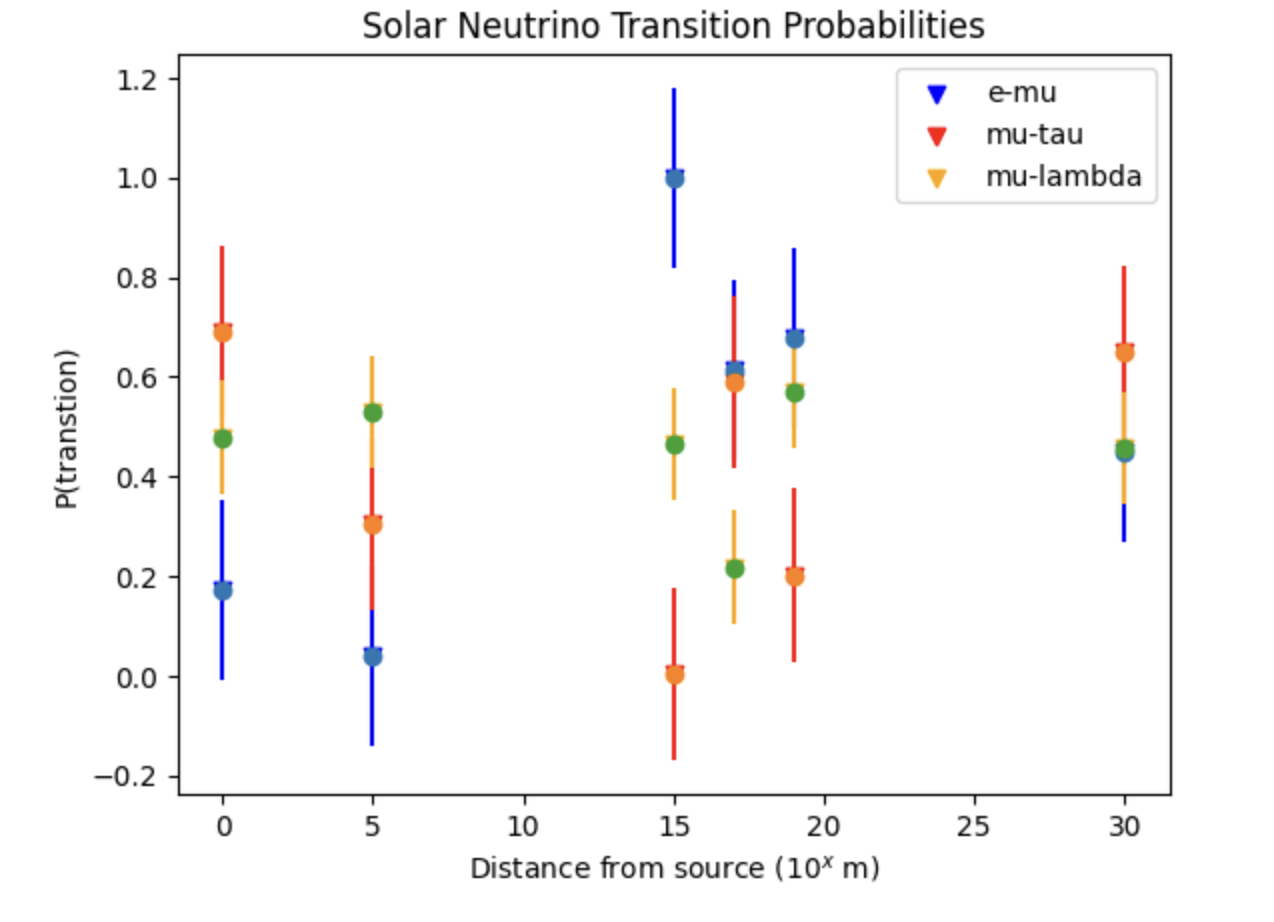}
\caption{Scatter plot depicting the comparison of probabilities between solar neutrino flavor transitions  $\nu_{e}\rightarrow\nu_{\mu}$, $\nu_{\mu}\rightarrow\nu_{\tau}$ and $\nu_{\tau}\rightarrow\nu_{\lambda}$.}
\label{fig: SOLAR}
\end{figure}

\begin{figure}[h!]
\centering
\includegraphics[width=0.45\textwidth]{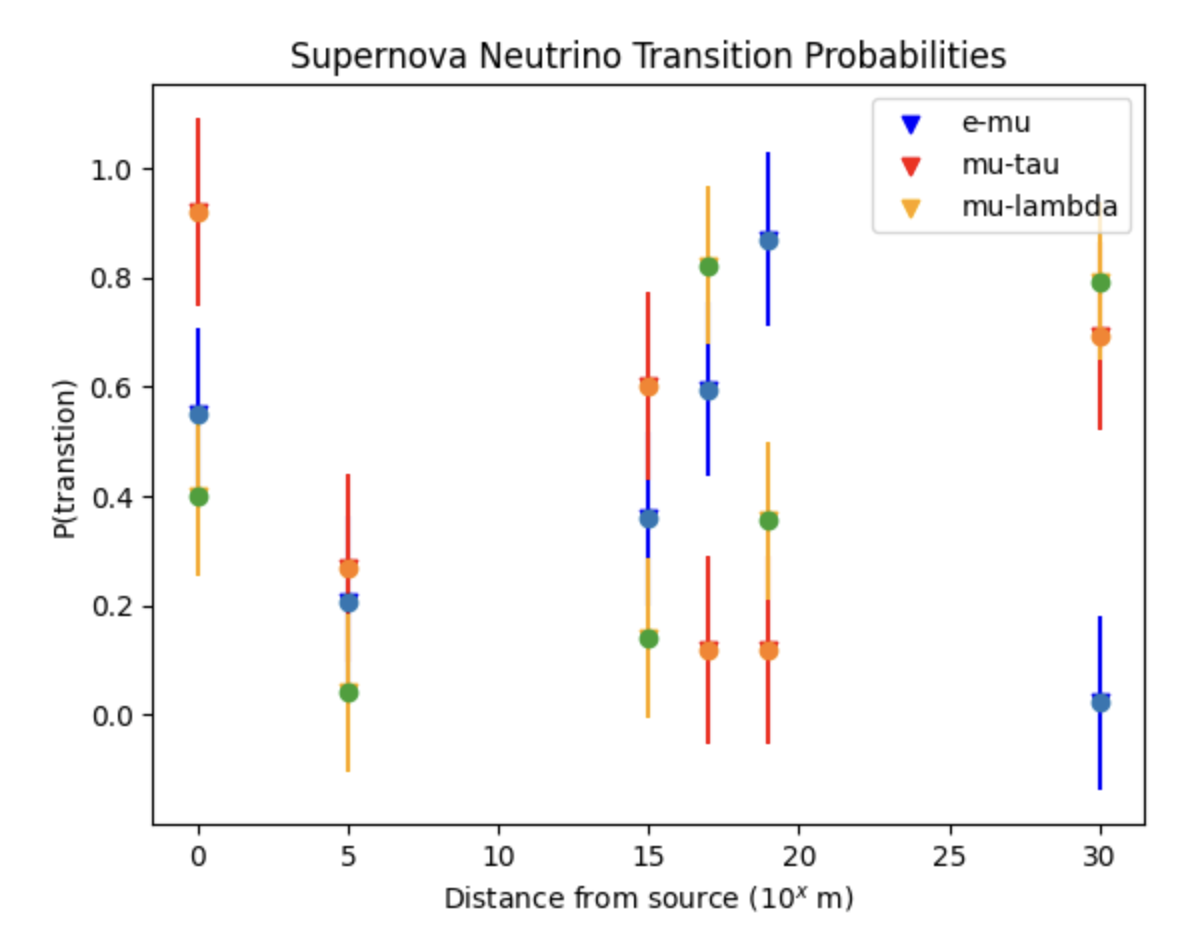}
\caption{Scatter plot depicting the comparison of probabilities between supernova neutrino flavor transitions  $\nu_{e}\rightarrow\nu_{\mu}$, $\nu_{\mu}\rightarrow\nu_{\tau}$ and $\nu_{\tau}\rightarrow\nu_{\lambda}$.}
\label{fig: SUPER}
\end{figure}

\begin{figure}[h!]
\centering
\includegraphics[width=0.45\textwidth]{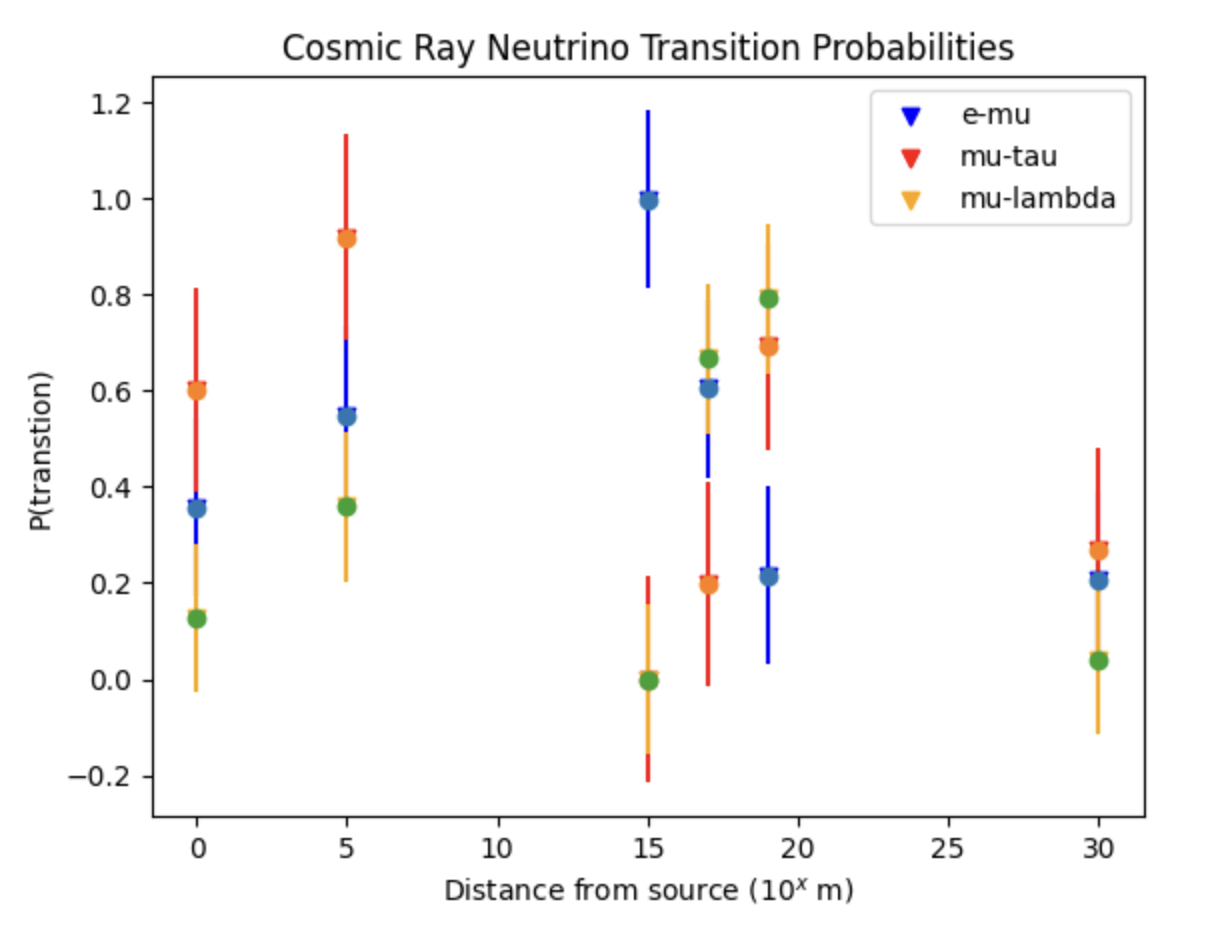}
\caption{Scatter plot depicting the comparison of probabilities between cosmic ray flavor transitions  $\nu_{e}\rightarrow\nu_{\mu}$, $\nu_{\mu}\rightarrow\nu_{\tau}$ and $\nu_{\tau}\rightarrow\nu_{\lambda}$.}
\label{fig: COSMIC}
\end{figure}

In Figures \ref{fig: SOLAR}-\ref{fig: COSMIC}, we show a comparison between the probabilities of the flavor transitions: $\nu_{e}\rightarrow\nu_\mu$, $\nu_{\mu}\rightarrow\nu_{\tau}$ and $\nu_{\tau}\rightarrow\nu_{\lambda}$ for solar, supernova and cosmic ray neutrinos, where $\lambda$ is any higher order flavor transition from the $4^{th}...n^{th}$ flavor states. We can see from the probability distribution, that the closer we are to the source, the lower the probability of a higher-order flavor transition, as expected. However, as the neutrino oscillates at greater distances, a pattern of fluctuating probabilities can be observed. We can infer from this that as the neutrino approaches $\sim10^{25}$ m from its source, the probabilities for all three flavors are all three approximations of each other, $\pm0.10$. With this understanding we can then follow the flavor transition by reverting backward through each flavor culminating at an equivalent point for both $\nu_{e}$ and $\nu_{\mu}$. From there, and as the neutrino oscillates further, we can then assume a probability for the first flavor as it approaches the probability for the $n^{th}$ flavor transition; in this case, the sterile neutrino. 
\par
From the data in the figures above, we can ascertain that at a certain point, the probability of a neutrino oscillating into a fourth flavor state after a certain distance approaches the probability that the neutrino is being found in its first flavor state. This further shows that each flavor can be thought of as a proper combination of all mass and flavor eigenstates oscillating back and forth between states. The probability of oscillating into a higher $n^{th}$ flavor is approximately the same as that of the neutrino oscillating backward through the previous states and beginning the cycle.
\par 
We can see in Figure \ref{fig: SOLAR}, as the solar neutrino has traveled $\sim10^{27}$ m from its source, the probability of it being found in the electron state is approximately the probability of it being found in its lambda state. From Figure \ref{fig: SUPER}, we can see the same equivalency; however with a supernova neutrino the distance is now $\sim10^{20}$ m. With higher-energy neutrinos, such as those in Figure \ref{fig: COSMIC}, we can see a unique symmetry between all three flavor of neutrino states. Here we can see where one flavor has a high probability, and the other two are at their lowest, and vice versa. 
\par
When we consider the neutrino's mean free path of $\sim105,700$ light years, or $10^{21}$ m from its source, we can see that this fits perfectly with the neutrino energy stemming from supernova neutrinos. However, when we focus on the lower-energy solar neutrinos, we find that given a slightly further distance of travel ($10^{27}$ m), we arrive at a near $0.5$ probability of detecting our neutrino in the first and fourth flavor states. An inference here tells us that the lower-energy neutrinos, given a longer distance of oscillation, have a higher probability of oscillating back through and begins the cycle over again. higher-energy levels tend to oscillate much faster, creating a uniquely consistent probability distribution for flavor transformations.

\begin{figure}
\centering
\includegraphics[scale=0.30]{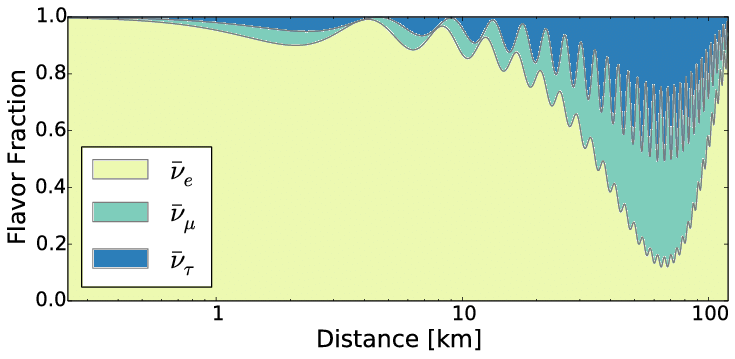}
\caption{The above image shows the fraction of detected neutrino flavors with their corresponding oscillation distance. Figure used with permission of author \cite{Vogel (2015)}.}
\label{fig: RENO}
\end{figure}

\pagebreak

\par
From Figure \ref{fig: RENO}, we can see the flavor fraction of detected neutrinos at the Reactor Experiment for Neutrino Oscillations (RENO) in South Korea \cite{Vogel (2015)}. When comparing this data to our above probabilities, and adjusting for unit of distance conversions, we find that for solar neutrinos at approximately $50$ km we deduce a probability of $\nu_{e}\sim0.985$, $\nu_{\mu}\sim0.928$ and $\nu_{\tau}\sim0.696$ and at $75$ km $\nu_{e}\sim0.473$, $\nu_{\mu}\sim0.419$ and $\nu_{\tau}\sim0.020$. In comparison with supernova neutrinos we find that at the same distances we have probabilities of $\nu_{e}\sim0.202$, $\nu_{\mu}\sim0.189$ and $\nu_{\tau}\sim0.0.33$ and at $75$ km $\nu_{e}\sim0.473$, $\nu_{\mu}\sim0.542$ and $\nu_{\tau}\sim0.763$. Based on the above data, and our previous calculations the probability for cosmic ray neutrinos remain on the order of $10^{-5}$ until it reaches an approximate distance of $\sim1\times10^{5}$ m or $100$ km which is also shown in the above fraction distribution. These observations were conducted at RENO in South Korea. 2023 began new observations at JUNO in Guangdong, Southern China. Observations should be concluded Late 2024-Early 2025. It will be of great interest to note any similarities and/or differences in the observations. 

\begin{figure}[h!]
\centering
\includegraphics[scale=0.45]{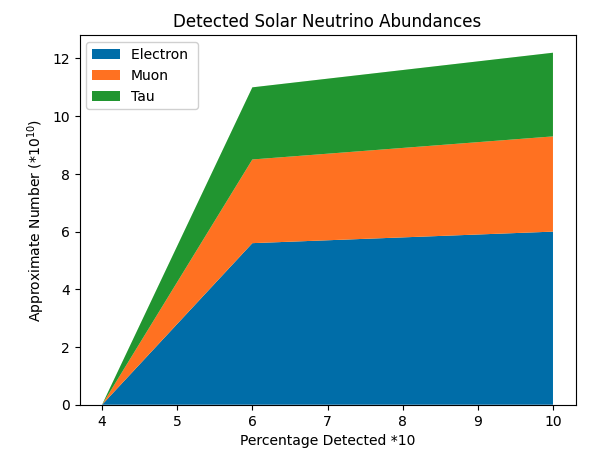}
\caption{The above image shows the fraction of predicted solar neutrino flavor detection given an oscillation distance of $1.5\times10^{11}$ m.}
\label{fig: SOLAR DEN}
\end{figure} 

The density chart described in Figure \ref{fig: SOLAR DEN} was calculated using a predicted flavor contribution to the solar neutrino flux detected on Earth. With an oscillation distance of $L_{E}=1.5\times10^{11}$ meters, we find our 
 changes to 
 neutrino 
 flavor abundance (number density)  
 with distance for each observed flavor
 to be $\frac{\Delta}{\Delta L} <n_{\nu_{e}}, n_{\nu_{\mu}}, n_{\nu_{\tau}}>\approx <n_{0}(1-P_{12}),n_{0}P_{12}(1-P_{23}),n_{0}P_{12}P_{23}(1-P_{3\lambda})>$ where, $n_{0}$ is the original number density of 
 electron 
 neutrinos emitted from proton-proton chains within the Sun which is approximately $6.0\times10^{10}\:\mathrm{cm}^{2}$ and $P_{mn}$ are the corresponding flavor transition probabilities.  
 \par
 The neutrino flux can be found by $\frac{2L_{Sun}}{28 MeV}\frac{1}{4\pi d^{2}}/\: \mathrm{cm}^{2}\approx 6.0\times10^{10}\: \mathrm{cm}^{2} \mathrm{s}^{-1}$. With an oscillation distance on the order of one AU, $10^{11}$ meters, we find the number of detected flavors we can expect on Earth from the Sun to be approximately $1.158\times 10^{11}$ neutrinos per square meter per second.

When comparing our probability density with the observed fraction density taken by RENO in South Korea \cite{Vogel (2015)}, we can see that our predicted flavor detection at much greater distances, closely resembles those observed by RENO at a tenth of the distance from the Sun. This tells us that within distances of approximately $100,000-1,500,000$ meters, our probabilities give us great confidence in our ability to predict flavor transitions over great distances. 

For comparative purposes, tables for both supernova neutrinos as well as those from cosmic rays were constructed using similar methods. However for the tables below, a supernova distance of $1.6\times 10^{21}\: \mathrm{m}$ was used and for the cosmic rays a shorter distance of $\sim 10^{11}\: \mathrm{m}$ was used for these particular densities. 

\begin{figure}[h]
\centering
\includegraphics[scale=0.40]{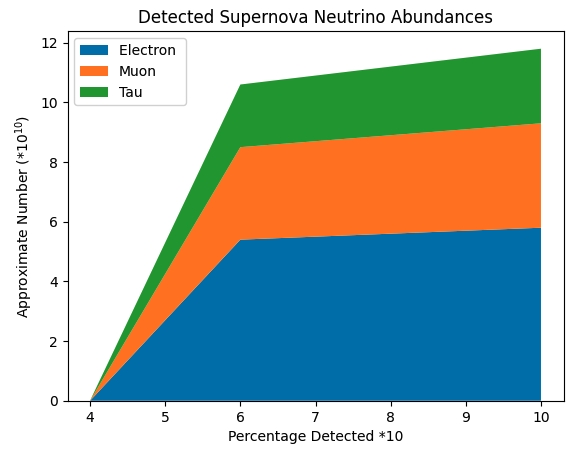}
\caption{The above image shows the fraction of predicted supernova neutrino flavor detection given an oscillation distance of $1.6\times10^{21}$ m.}
\end{figure}

\begin{figure}[h]
\centering
\includegraphics[scale=0.40]{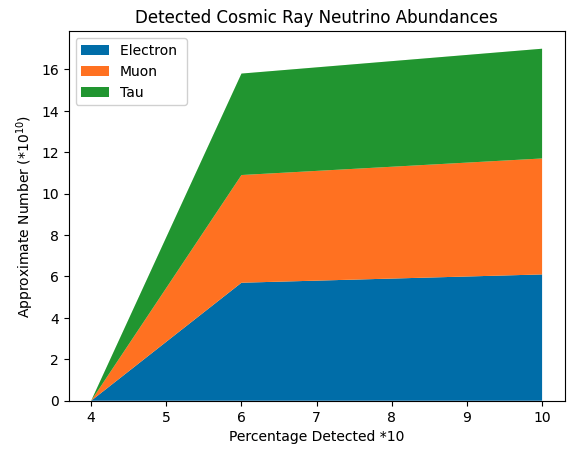}
\caption{The above image shows the fraction of predicted cosmic ray neutrino flavor detection given an oscillation distance of $\sim10^{11}$ m.}
\end{figure}

We found that for the supernova neutrinos, around $50\%$ should be detected as electron neutrinos while $\sim30\%$ to be detected as muon neutrinos and the final $20\%$ should be detected as tau neutrinos. Interestingly for the cosmic ray neutrinos, we found that approximately $36\%$ should be detected as electron neutrinos, $33\%$ as muon neutrinos and $31\%$ as tau neutrinos.

\section{Discussion}

In this paper we have shown that we can determine the probabilities of neutrino flavor states by first calculating the product of flavor eigenstate and then mass eigenstate probabilities. We found that the greater the distance from the source, the higher the probability of finding the neutrino in a different state than that in which it was produced. We found that for solar neutrinos, $10^{5}\: \mathrm{m}$ from the source, the neutrino has a $99.9\%$ chance of being detected in the state, $\nu_\mu$, and a $46.4\%$ chance of oscillating into a fourth flavor or simply reverting backwards. We find that at around $10^{27}\: \mathrm{m}$ this process starts over again. For supernova neutrinos, we find that at the same distances, we expect to find $\nu_\tau$ with a $60.2\%$ chance and $\nu_{e}$ with a $35.8\%$ chance. This time at around $10^{25}\: \mathrm{m}$ we find the process starting again. For the higher-energy cosmic rays, we find a similarity between the probabilities of that and the solar neutrinos. When comparing at the same distances, we have $99.9\%$ chance of detection in the state, $\nu_{e}$ and a $0.00\%$ chance of higher flavor transitions. Again, back to our distance of $10^{27}\: \mathrm{m}$, we find the process starts again.
\par
In repeating the same process, but for a distance of one AU unit, or $1.5\times 10^{11}\: \mathrm{m}$ we found that our neutrino flux should be comprised of approximately $50\%$ electron-neutrinos, $26.7\%$ muon-neutrinos and $23\%$ tau neutrinos. When we extend this distance a bit further for supernova we found approximately, $48\%$, $32\%$ and $20\%$ electron-, muon- and tau-neutrinos, respectively. When considering higher-energy cosmic rays at a distance of approximately $10^{11}\: \mathrm{m}$ we found a surprising split between $36\%$, $33\%$ and $31\%$ for the electron-, muon- and tau-neutrinos.
We have shown with minimal assumptions on well-established neutrino transition probability function that neutrino population statistics differ significantly among astrophysical sources. Both Solar and supernova neutrino populations detected on Earth are relatively  $\nu_e$-dominant, whereas higher-energy cosmic ray neutrinos that are candidates for Ice Cube, Super Kamiokande and future planned detectors have a relatively uniform flavor distribution. 

So far, only three flavor states have been detected, however we have shown that the likelihood of oscillating into a fourth flavor is equal to the oscillation back through the previous flavors, $\nu_{\tau}\rightarrow\nu_{\mu}\rightarrow\nu_{e}$. This raises an important question: since we know the flavors are linear combinations of the mass eigenstates, is it possible for a fourth flavor transition or would the neutrino's transformation cease at the $\nu_{\tau}$ and simply revert back to the first flavor state? 
\par
To put it more concisely, measuring in the flavor state prevents us from identifying the mass state, except as a linear combination of all possible masses. In future research, we may explore whether the assumption of non-locality suggests that mass eigenstates can also be represented as flavor eigenstates. We determined that when  examining the probabilities of detected states, we can make predictions based on the distance traveled since their creation. We discovered that a neutrino not only has a nearly certain probability of oscillating further but also of oscillating back into earlier states. This opens up opportunities for further research to focus on mass eigenstates and investigate if combinations of masses can replicate other flavors, and vice versa.

\textbf{Acknowledgements} \\
\indent The authors wish to thank Joseph A. Formaggio at MIT and Petr Vogel at Caltech for their permission to use Figure 1 and Figure 5, respectively.


\end{document}